\documentclass{aa}     
\usepackage{graphicx}

\newcommand{\Mpc}{$h^{-1}$\thinspace Mpc}

\newcommand{\Msun}{\mbox{M}_{\sun}}

\begin{document}

\title{Light-cone Simulations: Evolution of
dark matter haloes}

\author {P. Hein\"am\"aki\inst{1,2}, I. Suhhonenko\inst{1,2},
E. Saar\inst{1}, Maret Einasto\inst{1}, Jaan Einasto\inst{1},
Heidi Virtanen\inst{3}}

\authorrunning{P. Hein\"am\"aki et al.}

\offprints{M. Einasto }

\institute{ Tartu Observatory, EE-61602 T\~oravere, Estonia
\and
Tuorla Observatory, V\"ais\"al\"antie 20, Piikki\"o, Finland
\and
Division of Theoretical Physics, P.O.Box 64
FI-00014 University of Helsinki, Finland}

\date{ Received   2005 / Accepted ...  }

\titlerunning{LCS}

\abstract{We present a new fast method for simulating pencil-beam
type light-cones, using the MLAPM-code (Multi Level Adaptive
Particle Mesh) with light-cone additions.  We show that by a careful
choice of the light-cone orientation, it is possible to avoid extra
periodicities in the light-cone.  As an example, we apply the method
to simulate a 6 Gpc deep light-cone, create the dark matter halo
catalogue for the light-cone and study the evolution of haloes from
$z=6$ up to the present time. We determine the spatial density of
the haloes, their large-scale correlation function, and study the
evolution of the mass function. We find a surprisingly simple
relation for the dependence of halo maximum mass on redshift, and
apply it to derive redshift limits for bright quasars.

}

\maketitle

\section{Introduction}

The recent (completed and on-going) galaxy redshift surveys
(e.g.  LCRS, 2dFGRS, 6dF, SDSS) of our local universe have
enormously increased our knowledge of galaxies and of their large
scale distribution.  These large-scale structure data sets provide
also a powerful tool for breaking many of the parameter degeneracies
associated with the CMB data (Spergel et al. \cite{sper}).  Clusters
of galaxies, the largest virialised systems known, have a vital role
in understanding cosmological structure formation.  In particular,
following the evolution of clustering with redshift, we can put direct
constraints on models for the evolution of density perturbations
(Munshi et al. \cite{mun}).

However, it is impossible to study cosmological evolution, using only
such local surveys.  Thanks to recent developments in instrument
technology there are several new surveys designed to probe at
increasingly higher redshifts, in order to study the early evolution
of galaxies and their systems. We shall list a few of them.

The DEEP2 redshift survey is planned to study the evolution of
properties of galaxies and the evolution of the clustering of galaxies
from $z \sim 1.5$ to $z=0$, using the DEIMOS spectrograph on the 10 m
Keck II telescope.  The results will be compared with those of the
local surveys, e.g., as the LCRS (Coil et al. \cite{Coil}).
The ALHAMBRA photometric survey (Moles et al. \cite{moles05}) has similar
goals.
It will cover 4 square degrees of sky, and will find galaxies up to
redshifts $z\approx5$.

The VIRMOS galaxy redshift survey, using VIMOS on the VLT telescope,
is also designed to study the formation of galaxies and large-scale
structure over the redshift range ($0<z<5$), covering sixteen square
degrees of the sky in four separate fields (Virmos Consortium,
http://www.oamp.fr/virmos/).

GOODS (the Great Observatories Origins Deep Survey) aims to unite
extremely deep multi-wavelength observations (up to $z\approx 6$) both
from ground-based (VLT, Keck, Gemini, NOAO, Subaru) and space
telescopes: Hubble (HST Treasure program), SIRTF (Legacy Program), and
Chandra.  Observations cover two fields ($10'\times16'$), centered on
the Hubble Deep Field North and the Chandra Deep Field South.  The
primary goal of the GOODS program is to provide observational data for
tracing the mass assembly of galaxies throughout most of cosmic
history (Dickinson \& Giavalisco \cite{dic}).

The Galaxy Evolution Explorer (GALEX) satellite was launched in 2003
(Martin et al. \cite{m05a}). In combination with ground based optical
observations this satellite allows to study the star formation rate,
the galaxy luminosity function and other parameters of galaxy
evolution over the redshift range $0<z<5$, and even beyond this
limit (see http://www.galex.caltech.edu).

Several recent studies have revealed clustering of Ly alpha objects,
quasars and radio galaxies at very high redshifts (Ouchi et
al. \cite{ouch}, Venemans et al. \cite{ven}, Malhotra et
al. \cite{mal05}, Wang et al. \cite{wang05}, and
Stiavelli et al. \cite{stia05}).
For example, Ouchi et al. (\cite{ouch}) reported about the discovery
of primeval dense structures at redshift 6 containing Lyman alpha
emitters (LAEs). Ouchi et al. estimated that masses of these
structures - progenitors of present-day large scale structures are of
order of $10^{12} M_{\sun}$ - $10^{13} M_{\sun}$ and dimensions are
less than 10 \Mpc.

By these surveys, observations are reaching the scales where the
evolution of galaxies along the light-cone plays an essential role.
Comparison of deep surveys with our theoretical understanding of
structure formation and evolution has become even more topical and
thus simulations, or mock catalogs, have become an important tool for
the design of observational projects and for later data analysis.  A
properly constructed simulation can allow us to compare the
observations directly with modern models of structure formation, can
serve to test the data processing algorithms for biases, and to
quantify the impact of numerous observational selection effects (Yan
et al. \cite{yan}).

We believe that light-cone simulations mimic better the
observational datasets than conventional simulations at $z=0$, and
at different ``snapshots'' at fixed redshifts.  The light-cone
algorithm stores 'light-cone particles' on-the-fly, allowing us to
follow continuously the evolution of structure.  Although not
difficult in principle, light-cone simulations tax heavily computer
resources -- the volumes are too large to model them with necessary
mass resolution.  The best-known light-cone, ``the Hubble
simulation'' (Colberg et al. \cite{colberg}), was run on specialised
parallel shared-memory computers, and remains the only one with
publicly available data.

The goal of this paper is to present a new method to perform
light-cone simulations of deep pencil beams, typical for deep
surveys. Our method is rather efficient and allows to perform and
analyse the simulation on single workstations. We shall build a
light-cone from $z=0$ to $z=6$ in a ``concordance'' cosmological
model, and shall follow the evolution of dark matter (DM) haloes with
the cosmic epoch.  Visualizations of DM-haloes in the light-cone model
can be seen at Tartu Observatory web pages ({\tt
http://www.aai.ee/$\sim$maret/lc.html}).

When our project was almost finished we learned that a similar
light-cone project MoMaF has been realized by Blaizot et al.
(\cite{blaizot05}). Below we shall compare their approach with ours.

\section{The light-cone simulation}

\subsection{Method}

We use the Multi Level Adaptive Particle Mesh code (MLAPM, Knebe et
al. \cite{knebe}) with light-cone output.
The MLAPM code is adaptive, with sub-grids being adaptively formed in
regions where the density exceeds a specified threshold. There are
similar codes, as the ART code, Kravtsov et al. (\cite{kratso}); it is
essential that simulations are done with different codes, to
cross-check the results of the complex dynamics of the cosmic
structure formation.

We base the light-cone on periodic replicas of our simulation
cube. This could, of course, introduce extra periodicities in the
results, but we shall demonstrate that a clever choice of the
light-cone parameters will avoid those, at least for pencil-beams.  If
we choose the pencil-beam along one of the coordinate axes, we will
sample the same regions of the cube many times; but if the direction
of the beam is oblique, and the beam narrow, we will use different
regions of the cube along the beam.

In order to collect the light-cone, we find first the co-moving
light-cone radius and determine all the copies of the original
simulation cube, which intersect with the spherical surface of that
radius (cover it). For the simplest geometries (almost full-sky) that
is enough, but for a pencil-beam patch geometry we have to use three
additional checks:
\begin{enumerate}
\item find if the cube copy vertices lie inside the patch;
\item find if the patch edges intersect the copy's faces;
\item find if the patch planes intersect the copy's edges.
\end{enumerate}

If nothing this happens, the copy-cube does not include the
light-cone. This cube list may seem to be an unnecessary complication,
but it will speed up light-cone processing for the most interesting
deep and narrow light-cones.

When the cube list is completed, we check for all particles their
positions in all the candidate cubes, and if the (copy) particle
has crossed the light-cone between the preceding $z$ and the present $z$,
we find the crossing redshift and coordinates by linear interpolation.
Then we check for the patch geometry, if necessary, and write the
(copy)-particle data.

Detailed instructions on generation of the light-cone simulations can
be found in the MLAPM user guide
(http://www.aip.de/People/AKnebe/MLAPM).

The MoMaF project (Blaizot et al. \cite{blaizot05}) takes a slightly
different approach, saving first a large number of snapshots and
using these later to build the light-cone. When assembling the
light-cone, they randomly rotate the snapshots, in order to
eliminate the influence of periodicities. As this destroys the
continuity of the structure, we prefer not to scramble our cubes. In
fact, MoMaF allows such an approach too, for pencil-beam surveys,
although Blaizot et al. (\cite{blaizot05}) devote most of their
attention to the scrambled case.

\subsection{Application}

We simulate a pencil-beam mock survey, similar to the contemporary
observational surveys. We use a simulation with $256^3$ dark matter
particles in a $256^3$ grid. We follow the evolution from $z=6$ to
present time.  Our simulation covers $2^{\circ} \times 0.5^{\circ}$ in
the sky.

We use a flat cosmological model with the parameters derived by the
WMAP microwave background anisotropy experiment (Bennett et
al. \cite{bennet}): the dark matter density $\Omega_m=0.226$, the
baryonic density $\Omega_b=0.044$, the vacuum energy density
(cosmological constant) $\Omega_{\Lambda}=0.73$, the Hubble constant
$h=0.71$ (here and throughout this paper $h$ is the present-day Hubble
constant in units of 100 km s$^{-1}$ Mpc$^{-1}$) and the rms mass
density fluctuation parameter $\sigma_8=0.84$.

The initial data for our simulation was generated in a cube of
200~\Mpc\ co-moving size, for $z=30$.  Thus, each particle has a mass
of $3.57\cdot 10^{10} h^{-1}\Msun$. The transfer function for our
model was computed using the COSMIC code by E.~Bertschinger ({\tt
http://arcturus.mit.edu/cosmics/}).

\begin{figure*}[ht]
\centering
\resizebox{1.57\columnwidth}{!}{\includegraphics*{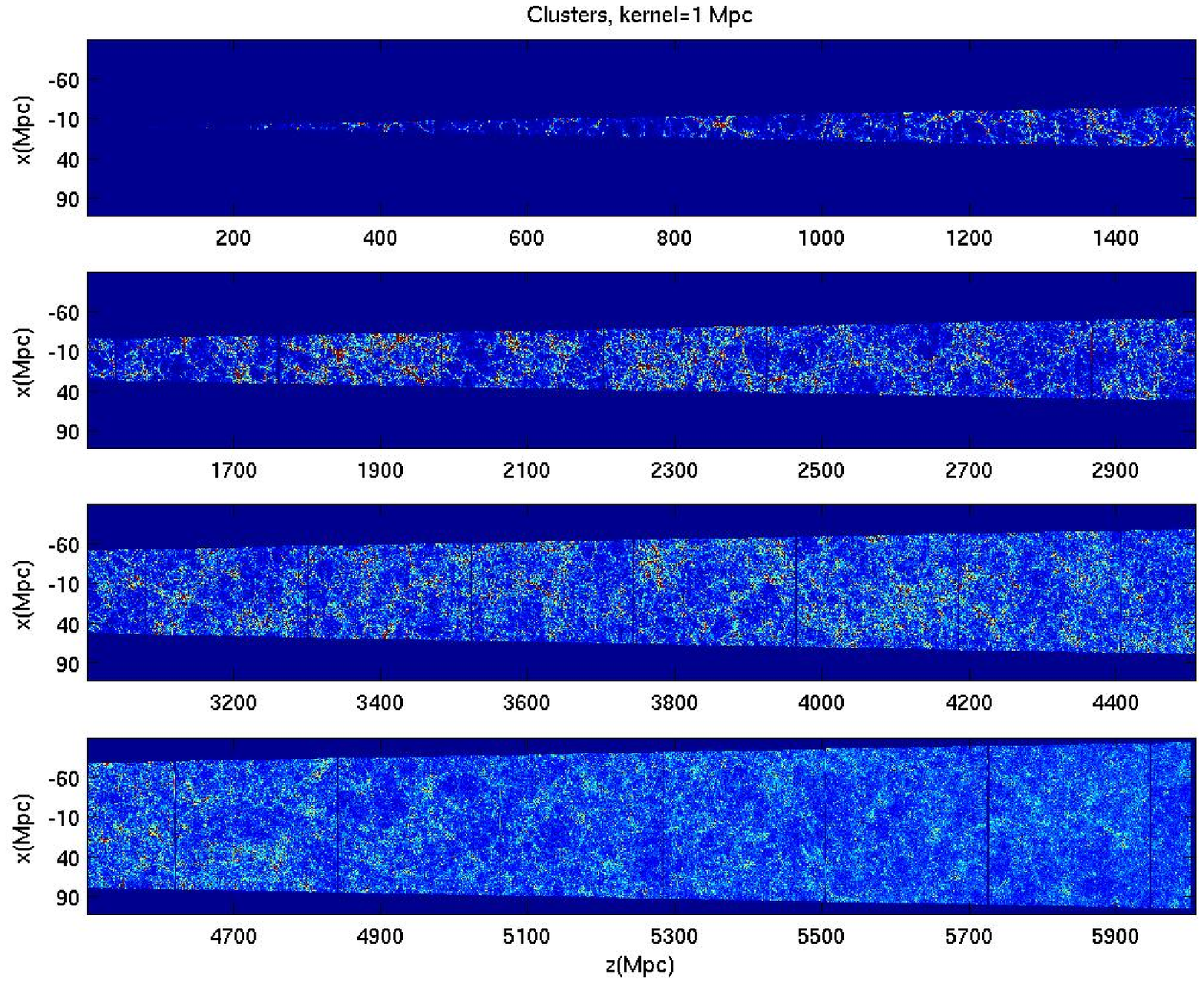}}
\resizebox{1.57\columnwidth}{!}{\includegraphics*{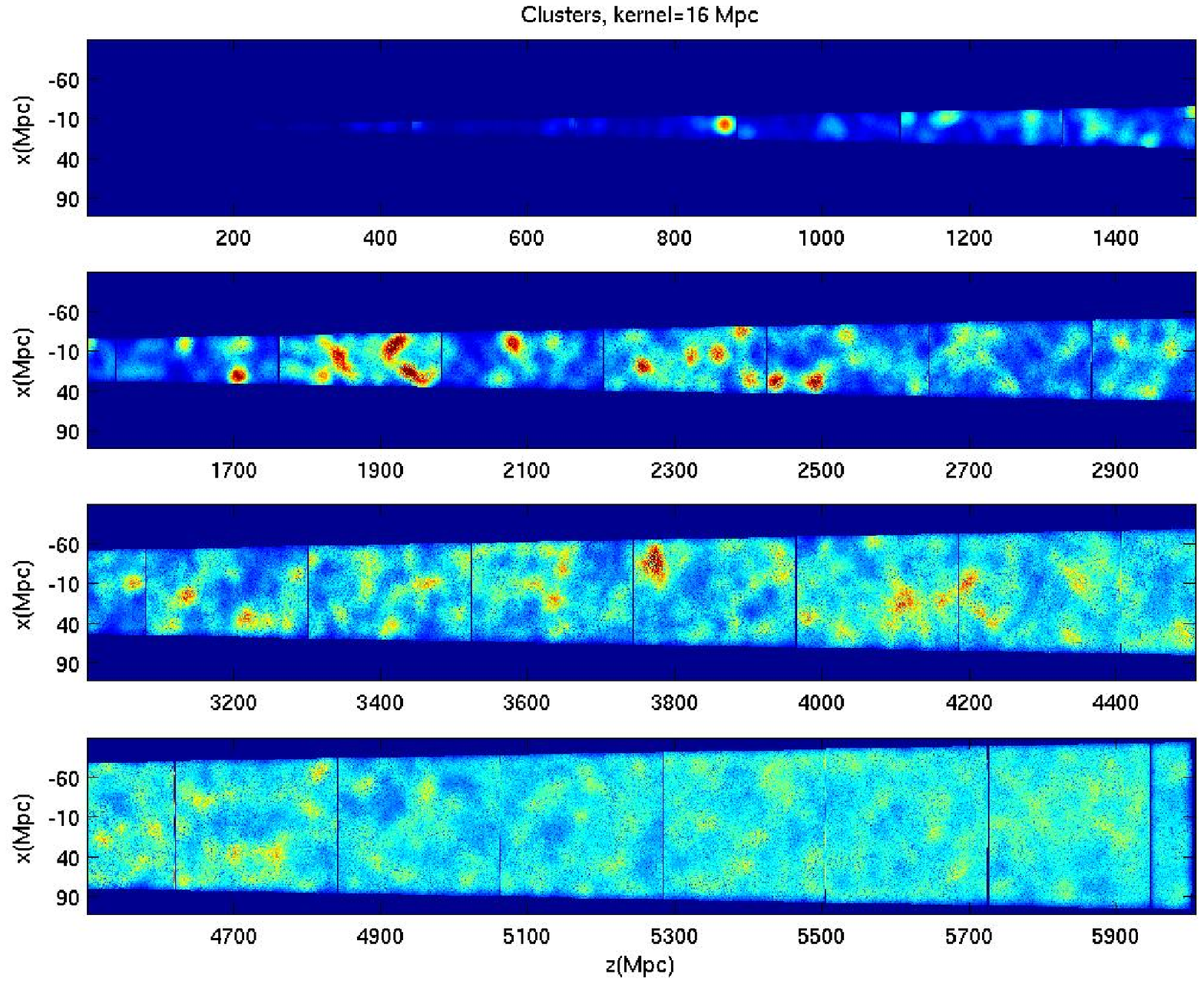}}
\caption{The density field of the 6 Gpc deep dark matter
light-cone simulation smoothed with a
$r = 1.0$~\Mpc\ kernel (upper panels), and
with a $r = 16.0$~\Mpc\ kernel (lower panels).}
\label{fig:1}
\end{figure*}

Figure \ref{fig:1} shows the projected (dark-matter) density fields
of our light-cone simulation from $z=0$ to $z=6$ (for the
cosmological model chosen $z=6$ corresponds to 5981.42 \Mpc.  The
density was calculated, using an Epanechnikov kernel
($h(x)\sim(1-x^2/r^2)$ of radii $r$ 1.0~\Mpc, 16.0~\Mpc\ - i.e., for
the characteristic scales of clusters and superclusters of galaxies.

The full amount of light-cone data (positions and velocities of dark matter
particles) can have enormous. The evident way to compress
the data is to extract catalogues of dark matter haloes; this are the
objects that can be associated with observed galaxies and clusters of
galaxies. 

As the amount of data is large, we used the simplest algorithm
(friends-of-friends, FOF) to identify dark matter haloes. The
simulations reported in this paper were run in early 2004; now we
would have used a new feature of MLAPM, its halo finder (see Gill
et al. \cite{gill04}) that uses the basic adaptive density
tree of the simulation, and outputs haloes at all required
snapshots. It should not be difficult to adapt that to output
light-cone halo catalogs on the fly; this remains to be done.  Deep
light-cones generate huge amounts of data, and the only practical way
to handle this is to keep only the haloes.

FOF uses a linking length $b$ to collect particles in groups with
spacing closer than $b$ times the mean inter-particle spacing. The
linking length is frequently determined from the virialisation density
$\rho_v$ (see Bryan and Norman \cite{bry}), obtained from the solution
for the collapse of a spherical top-hat perturbation.  The density
$\Delta_c$ dependence on background cosmology via the matter density
parameter
\[
\Omega_M(z)=\Omega_0(1+z)^3/E(z)^2,
\]
where
\[
E^2=\Omega_0(1+z)^3 +\Omega_R(1+z)^3+\Omega_{\Lambda}.
\]
Bryan and Norman \cite{bry} give an approximate formula
for that:
\[
\Delta_c=\rho_v/\rho_{\mbox{crit}}=18 \pi^2+82x-39x^2,
\]
where $x=\Omega_M(z)-1$ and $\rho_{\mbox{crit}}$ is the density in the
Einstein-deSitter model ($\Omega_{\Lambda}=0, \Omega_M=1$).  The
actual mean density is $\Omega(z)\rho_{\mbox{crit}}$.  Thus the
linking length that would select virialised haloes at a given redshift
$z$ can be written as $b=(\Omega_M(z)\Delta_c)^{\frac{1}{3}}$.

The assumption that objects populate only virialised haloes could be a
bit extreme. We normalised the linking length by its present value,
comparing the amplitude of the simulated halo mass function (at it's
massive end) with an observational mass function of galaxy groups of
the Las Campanas Loose Groups of Galaxies, hereafter LCLG,
Hein\"am\"aki et al. (\cite{hei03}).  In order to obtain this mass
function, we ran the MLAPM code in the snapshot mode, with the same
initial data we used for the light-cone simulation.

Using this normalisation, we chose the present epoch ($z=0$) linking
length as $b=0.23$ (in units of the mean particle separation), which
corresponds to the matter density contrast $\delta n/n = 80$. This is
somewhat lower than the virial density, but not much, and delineates
well galaxy groups, as seen in the LCLG Catalog (Tucker et al.
\cite{tucker}).

\begin{figure}[h]
\centering
\resizebox{1.0\columnwidth}{!}{\includegraphics[angle=-90]{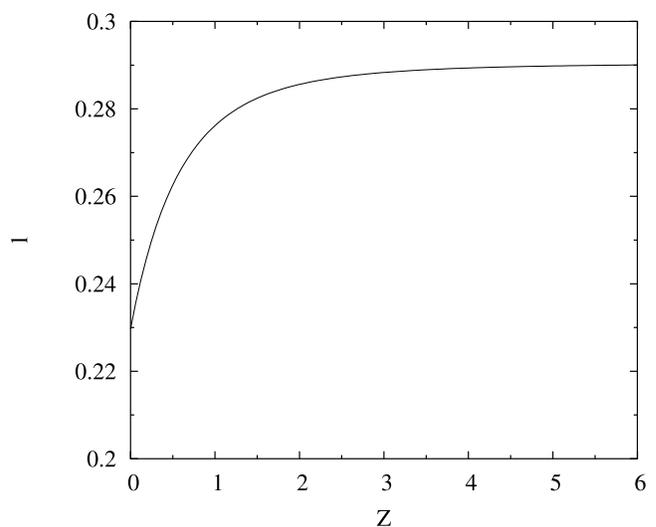}}\\
\caption {The FOF linking length as a function of redshift,
for the light-cone simulation.}
\label{fig:4}
\end{figure}

Figure ~\ref{fig:4} shows how the linking length we used changes with
redshift. Due to the accelerated expansion at later epochs the haloes
need higher density contrasts (and smaller linking lengths) for
virialisation.

We extracted several populations of dark matter haloes from our
light-cone: haloes with 2 or more particles
($M_h \ge 7.1\cdot 10^{10}h^{-1}\Msun$), haloes with more than 20
particles ($M_h \ge 7.1\cdot10^{11}h^{-1}\Msun$), and a conservative
sample of haloes with more than 100 dark matter particles
($M_h\ge3.6\cdot10^{12}h^{-1}\Msun$), We cleaned all the group
catalogues, using the virial equilibrium condition
$E_r=E_{kin}/|E_{pot}|<0.5$ ($E_{pot}$ is the potential energy and
$E_{kin}$ -- the kinetic energy of a group) for groups to be included
in our final group catalogue.

The number of small haloes is considerably reduced by the virial
condition, which is natural -- many of such haloes are random
encounters. From the $n_{\mbox{min}}=20$ halo sample, 6.5\% of haloes
are rejected, and only a few haloes from the rich halo sample do not
satisfy the virial condition.  This indicates that the linking length
we have chosen does not pick up too many non-virialised objects.  The
final full halo sample includes about 600,000 haloes, the intermediate
sample has 14016 haloes and the sample of rich haloes includes 4088
haloes.

\section{Checking for periodicities}

Assembling a light-cone from periodic replicas of a simulation cube
will lead to several statistical biases. These are thoroughly
discussed in Blaizot et al. (\cite{blaizot05}). The main bias comes
from scrambling the snapshots; as we do not use scrambling, we
bypass this bias.

Another bias is caused by the finite volume of the simulation cube
and can be corrected for; Blaizot et al. (\cite{blaizot05}) show how
to do that. We can also, in principle, to modify the light-cone by
linear large-scale modes; this would allow us to model better
correlation functions and number counts.

\begin{figure}
\centering
\resizebox{\columnwidth}{!}{\includegraphics*{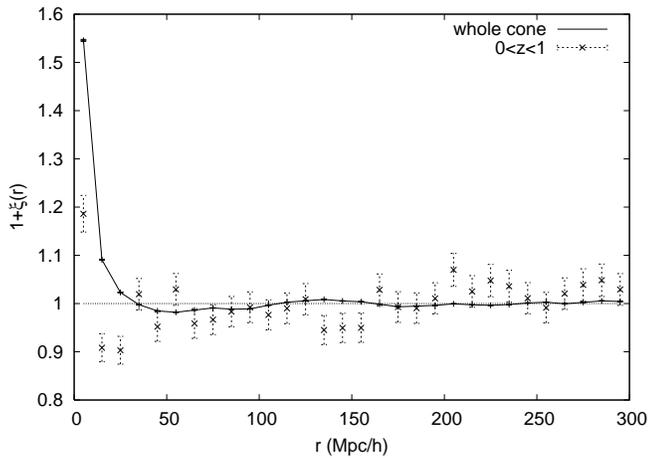}}
\caption{The large-scale
correlation function of DM haloes in the light-cone.
The solid line with error-bars shows the correlation function
for the whole sample, dotted bars show the correlation
function and its 1-$\sigma$ deviation for the redshift
interval $z\in[0,1]$.
\label{fig:corr}}
\end{figure}

There is one more effect to worry about -- as our light-cone is
composed of periodic replicas of a much smaller simulation cube, it
could have periodicities at scales of about the cube size. The
easiest way to check for that is to look at the two-point
correlation function of the light-cone. The total number of
light-cone mass points is very large; in order to calculate the
correlation function we chose the full halo sample, and diluted it
randomly, by a factor of six, to keep the $O(N^2)$ calculation time
within reasonable limits. The number of haloes used was about
10,000. We are worried about large-scale correlation, thus we chose
the Landy-Szalay estimator, which is one of the best estimator for
large scales (see, e.g., Mart{\'\i}nez \& Saar \cite{martsaar}).
The rms error of the estimator was taken to be twice the Poisson
error. We show this correlation function in Fig.~\ref{fig:corr}.

As we see, no periodicity at scales around 200~\Mpc\ can be
seen. The dip in the correlation function at $\sim$50 \Mpc\ and the
maximum around $\sim$120 \Mpc\ are real, and are seen both in
simulations and in the observed large-scale correlations (e.g., for
Abell clusters, Einasto et al. \cite{ein97}, see also Einasto et
al. \cite{ein94} and Tago et al.  \cite{tago02}).  A slight indication
of periodicity could be seen in the closest redshift interval
($z=0\dots1$, shown with dotted error-bars). However, even there
$\xi(r)=0$ for distances $r>30$\Mpc, within $2\sigma$ limits
(the volume and the number of haloes are small, and the estimate is
rather uncertain).  For other redshift intervals the correlation
functions practically coincide with the mean for the whole sample.

This can be explained by the good choice of the light-cone
geometry. An oblique direction of the light-cone with respect the
symmetry axes of the cube will cause the light-cone to sample
different regions of the simulation cube in its different replicas. We
can quantify this by counting the number of times each cube volume
element is included in the light-cone. In order to estimate that, we
choose a simple algorithm -- we populate the light-cone with a Poisson
point process, fold it back onto the original simulation cube, and
find the resulting one-point density distribution in this cube. When
properly normalised, this distribution approximates the volume
multiplicity distribution.

\begin{figure}
\centering
\resizebox{\columnwidth}{!}{\includegraphics*{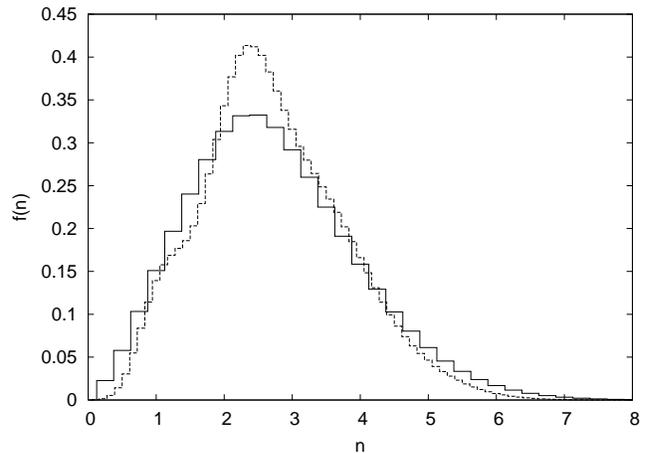}}
\caption{The volume multiplicity histograms of  the light-cone.
\label{fig:volmult}}
\end{figure}

The volume multiplicity distribution for our light-cone is shown in
Fig.~\ref{fig:volmult}. The two different histograms show the results
for two different oversampling factors (the ratio of the density of
the Poisson process to the mean light-cone density) -- the higher this
factor, the better the estimate, and the larger the computer time.

We see that most of the volume of the simulation cube is used only
twice in our light-cone. The worst choice of the light-cone
direction, along an axis of the cube, would give a factor close to
30 (the depth of the light-cone divided by the cube size). We can
use the mean of the histogram or its rms error as the
figure-of-merit $M$ for the volume use; maybe the best choice would
be to use a a combination of these:
\[
M^2=\sigma^2_m+\bar{m}^2,
\]
where $m$ is the volume multiplicity, and $\bar{m}$ its mean value.
Thus, the figure-of-merit for an ideal case (e.g., the Hubble volume
$\pi/8$ sterradian light-cone, Colberg et al. \cite{colberg}) would
be 1; for our light-cone it is 2.9 (but could be over 30). Another
relevant characteristic could be the distribution of distances
between point replicas. A rough estimate of the mean replica
distance for our light-cone is 6000/3=2000 \Mpc; this is why we do
not see periodicities in the correlation function.

The light-cone figure-of-merit can be found before simulations, and
can be used to select the best orientation parameters for a given
light-cone geometry (its size and the size of the simulation cube).
Its value will also characterise the possible periodicities in the
light-cone. We plan to include the figure-of-merit code in the tools
section of the MLAPM code package.

\section{Evolution: redshift dependence of the dark matter haloes}

\subsection{Spatial density of haloes}

Light-cones as ours can be used for many purposes. As this paper is
mainly meant to present the light-cone construction method, we shall
give only a few examples below.

First, as our light-cone is pretty deep, we can follow the formation
of first massive haloes. Smaller haloes form earlier yet, and their
study would require a deeper light-cone.

\begin{figure}
\resizebox{\columnwidth}{!}{\includegraphics*{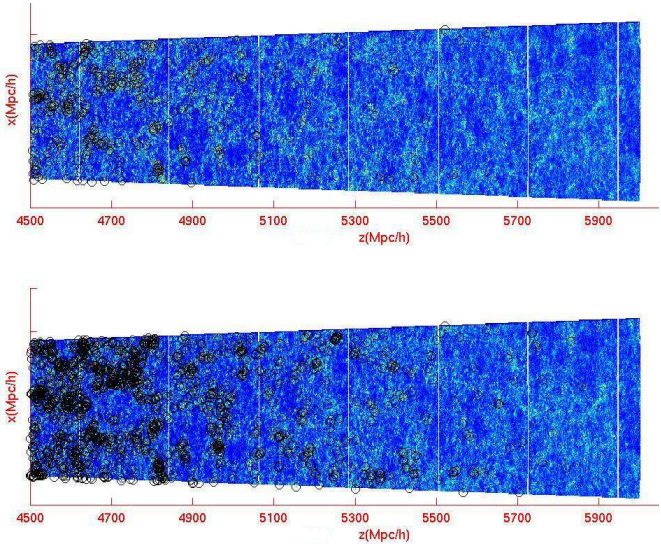}}\\
\caption{The density field of the simulation. Halos are marked with
black circles.  The figure shows the most distant part of the
light-cone, from 4 to 6 Gpc. The upper panel shows rich haloes ($M\ge
3.6\cdot 10^{12}h^{-1}\Msun$), and the lower panel shows also the
intermediate-mass haloes ($M\ge 7.1\cdot 10^{11}h^{-1}\Msun$).
\label{fig:5}}
\end{figure}

Figure ~\ref{fig:5} illustrates the appearance of the first dark
matter haloes. In panel (a), DM haloes with masses $M\ge 3.6 \cdot
10^{12}\Msun$ are shown.  Halos, placed on the density field map
(smoothed with 1 \Mpc\ kernel), are marked with black circles. As we
see, the very first massive haloes appear at about D=5600 \Mpc\
(z=4.97).  Panel (b) shows shows haloes of mass $M\ge 7.1 \cdot
10^{11}\Msun$. In this case the very first haloes appear already at
$D=5900$ \Mpc\ ($z= 5.62$).  Smaller haloes exist in our light-cone from
its start, $z=6$ already.

\begin{figure}
\centering
\resizebox{\columnwidth}{!}{\includegraphics*{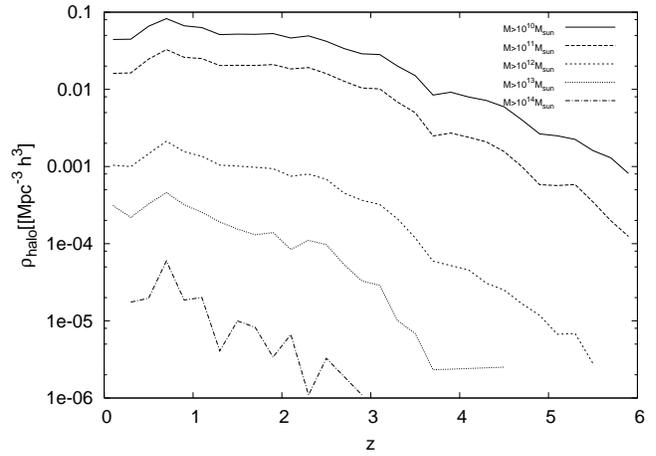}}
\caption{Spatial density of the haloes as a function of redshift for
five different mass limits.
\label{fig:6}}
\end{figure}

The spatial densities $\rho_{halo}$ of all our DM haloes as a function
of redshift (the number of haloes per co-moving volume at redshift bins
of width 0.2) for different halo mass limits are shown in Figure
~\ref{fig:6}.  Each line indicates different halo mass limits, from
$10^{10} \Msun$ (at the top) to $10^{14}\Msun$ (at the bottom of the
figure).  As can be expected, rich haloes appear later than poor ones.
Haloes with masses smaller than $10^{12}\Msun$ exist in our light-cone
from the beginning.  Overall density evolution of haloes of the two
smaller mass ranges is the same; in the redshift range $z=5.7\dots 4$
the spatial density of haloes increases about an order of magnitude.
More rapid increase occurs in the redshift interval $z=3.7\dots
3$. These transitions are also seen in the density behaviour of haloes
of masses $M\ge 10^{13}\Msun$.

The halo density reaches its maximum at $\sim z=0.9$, after that it
decreases. It could be caused by merging of smaller haloes, but could
also be a statistical fluctuation, due to the small volume of our
pencil beam at redshifts less than one.

The first haloes of mass $M \ge 10^{14}\Msun$ appear at $z \sim 3$.
The large variation in their density evolution is due to rareness of
such objects and to a small volume of the sample.  There is also two
overall slopes seen in mass scales less then $10^{14}\Msun$. We also
notice that density evolution on almost all mass scales (excepting the
the largest mass interval) between z=5.5 and z=3 is more rapid than
the later density evolution, between z=3 to z=1. The reason for that
could be gradual merging of smaller haloes to form bigger systems.

\subsection{Halo mass history}

\begin{figure}
\resizebox{\columnwidth}{!}{\includegraphics*{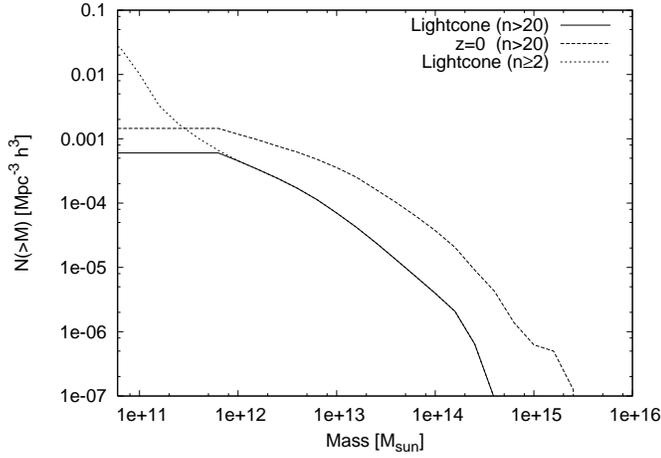}}\\
\caption{The halo mass function for the light-cone
(intermediate sample, $n\ge 20$) (solid line) and for a conventional
simulation for the present time (dashed). The dotted line
shows the light-cone halo mass function for all haloes.}
\label{fig:8}
\end{figure}

Figure ~\ref{fig:8} shows the cumulative mass function (MF) for dark
matter haloes, defined as the number density of haloes above a given
mass $M$, \mbox{$n(>M)$}.  The solid line shows the mass function for
the intermediate mass light-cone haloes ($n\ge20$).  The dotted line
shows the MF for a similar halo sample, obtained in our MLAPM
simulation, for the final snapshot (a 200 \Mpc\ box at $z=0$). The
mass functions differ by an order of magnitude at $10^{11}\Msun$, due
to the smaller overall halo density in the deep light-cone. Another
difference can be seen at the high mass end of the MF-s -- the $z=0$
MF extends to larger masses.  This is due to the narrowness of the
pencil beam, that occupies only about $10^{-4}$ of the simulation box
at $z=0$.  Statistics of such surveys should be taken with caution, at
least for redshifts less than $z\approx 1$. Without this cosmic
variance effect, the light-cone MF should follow the snapshot MF
throughout the whole mass range.  The dotted line shows the light-cone
mass function for all halo masses; this grows steeply, showing that
small haloes dominate in number, especially at early epochs, when
massive haloes did not exist yet.

\begin{figure}
\resizebox{\columnwidth}{!}{\includegraphics*{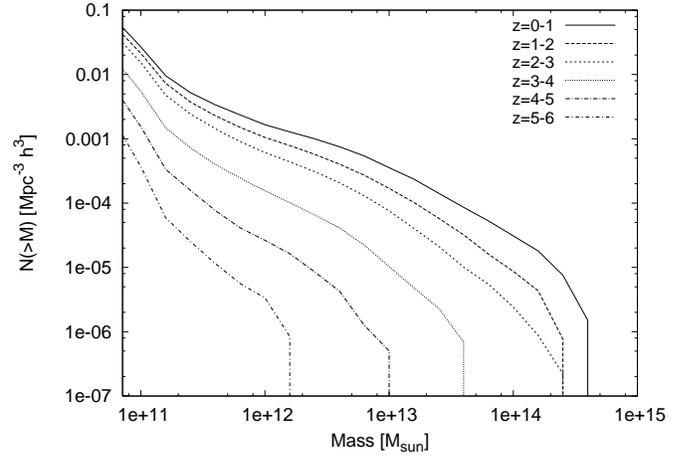}}\\
\caption{Halo mass functions for  different redshift intervals.}
\label{fig:9}
\end{figure}

The light-cone can be also used to study the evolution of the halo mass
function. We plot this function for different redshift intervals in
Fig.~\ref{fig:9}.  The overall shape of the mass functions between
$z=0$ to $z=1$ and $z=1$ to$z=2$ is quite similar to the mass function
of the whole sample, for the mass interval $M\ge 10^{12}\Msun$ (see
Fig.~\ref{fig:8}). The mass functions steepen with redshift, showing
the dominance of small haloes at large redshifts.  The characteristic
halo masses decrease with redshift, with the maximum halo mass at the
largest redshift interval $z=5\dots6$ being $1.5\cdot 10^{12}\Msun$.

\begin{figure}
\resizebox{\columnwidth}{!}{\includegraphics*{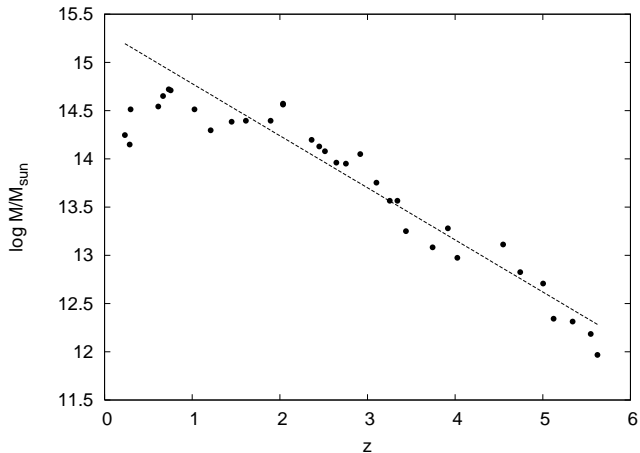}}\\
\caption{Masses of the most massive haloes for different redshifts}
\label{fig:7}
\end{figure}

The most easier objects to observe at any epoch are the most luminous
(most massive) ones.  In Fig.~\ref{fig:7} we plotted the maximum halo
mass for a given redshift.  This figure shows that the maximum masses
increase exponentially with decreasing redshift, down to the redshift
$\sim 0.9$, where the cosmic variance limit appears. This interesting
trend, shown in the figure, can be approximated by an intriguingly
simple formula
\begin{equation}
\label{eq:maxmass}
 M_{\mbox{max}}(z)\approx 10^{15}\Msun 10^{-0.5z}\approx 10^{15}\Msun e^{-z}.
\end{equation}
This relation is shown by a dashed line in the figure.

\section{Early haloes and black holes}

As an example of an application of light-cone models, we discuss an
interesting possibility that masses of supermassive black holes (BH)
may correlate with masses of dark matter (DM) haloes.

Using a sample of 37 local galaxies Ferrarese (\cite{fer}) obtained a
relation between the masses of BH and circular velocities of galaxy
disks, similar to the well-known relation between the` BH mass and the
bulge velocity dispersion (e.g. Magorrian et al. \cite{mag},
$M_{BH}/M_{bulge}=3\cdot 10^{-3}$).  Using Bullock et al.'s
(\cite{bul}) simulations (their result $M_{DM}=2.7\cdot
10^{12}(v_{vir}/200~km ~s^{-1})^3\Msun$ agrees well with our results)
together with Ferrarese's observational results, the best fit for the
ratio between $M_{BH}$ and $M_{DM}$ can be written as
\begin{equation}
\label{eq:massratio}
M_{BH}/10^8\Msun\sim 0.10(M_{DM}/10^{12}\Msun)^{1.65}
\end{equation}
(see figure 5 in Ferrarese's paper).  Ferrarese pointed out (see the
references therein) that the critical DM mass for the BH formation
could be about $5\cdot 10^{11}\Msun$, while less massive haloes might
not offer favorable conditions for BH formation. According to the
previous equation, this limits the BH masses by $\sim 10^6\Msun$ in
such DM haloes.  Ferrarese concluded also, that according to the
results by Haehnelt et al. (\cite{hahn}) only haloes with masses $M\ge
10^{12}\Msun$ may host a black hole with $M_{BH}\ge 10^6\Msun$.  In
our light-cone the spatial density of such host haloes is less than $2
\cdot 10^{-7}\mbox{Mpc}^{-3}h^3$ at redshifts larger than $z=5.5$.

If we assume that Ferrarese's $M_{BH}/M_{DMhalo}$ relation also holds
at high redshifts, and combine it with our maximum halo mass relation
(\ref{eq:maxmass}), we obtain a simple relation for the BH masses
likely to reside in the most massive dark matter haloes as a function
of redshift:

\begin{equation}
\label{eq:bhmass}
\frac{M_{BH}}{10^{12}\Msun}(z)=10^{-0.83z}.
\end{equation}

McLure \& Dunlop (\cite{mc}) concluded, based on virial BH mass
estimates from the SDSS first release and for 12698 quasars, that
quasars' SBH (super massive black hole) masses lie between $\simeq
10^7\Msun$ and $\simeq 10^9\Msun$.  If we substitute these limits in
equation (\ref{eq:bhmass}), we find that the most prominent quasars
exist between the redshifts $\sim 6$ to $\sim 2$. Haloes massive
enough to form SBH and hence quasars, either did not exist before the
redshift $z\approx 6$, or they were very rare, with the number density
$n_{h}<10^{-7}Mpc^{-3}h^3$.

\section{Conclusions and perspectives}

We have presented a new method for simulating pencil-beam type
light-cones, using periodic replicas of a base $N$-body cube.  Such
light-cones are typical for extremely deep optical surveys, either
underway or in the planning stage.  We have shown that by a careful
choice of the light-cone parameters, it is possible to avoid extra
periodicities in the light-cone.

We have simulated a deep (up to $z=6$) light-cone, generated the dark
matter halo catalogue, and studied its properties. We find that early
light-cone is dominated by small haloes, and the maximum halo mass can
be clearly traced throughout all the epochs.

We find a simple approximation for the dependence of maximum dark
matter halo mass on redshift, and use it to explain the redshift
limits of the quasar distribution.

In summary, the algorithm we use to simulate light-cones is
lightweight and fast. The code is in the public domain, included in
the MLAPM package. Nevertheless, the main bottleneck of any
light-cone model is the vast amount of data, proportional to $D^3$
($D$ is the depth of the light-cone). The only way to overcome this
is to discard most of the hard-obtained simulation data, the
positions and velocities of dark matter particles along the
light-cone, and to store only the data on dark matter haloes,
creating and analysing them on-the-fly. This could happen soon, as
our base MLAPM code already includes the MHF halo finder, which
outputs halo catalogues for fixed-time snapshots.
Such halo catalogs can be directly used to
predict the SZ effect from early galaxy clusters
for the Planck mission. Moreover, as gravitational
lensing directly traces the total matter density in the universe,
our light-cone catalogs are good tools to study the weak lensing
of the light emitted by distant quasars, and of the CMB.
These are the lines of our present work.

Another important problem in simulating light-cones to compare with
observational data is populating dark matter haloes with galaxies,
and assigning the galaxies the features an observer sees. In fact,
the MoMaF project (Blaizot et al. \cite{blaizot05}) already gives
galaxy populations in light-cones, using semi-analytic recipes. We
plan to include such recipes in our code, also.

\begin{acknowledgements}

We thank Alexander Knebe for for his support and interest in the
work; Mirt Gramann for stimulating discussions, and Vicent
Mart{\'\i}nez for valuable suggestions. The present study was
supported by the Estonian Science Foundation grants 4695 and 6104,
by the Estonian Ministry for Education and Science grant TO
0060058S98, by the University of Valencia through a visiting
professorship for Enn Saar and by the Spanish MCyT project
AYA2003-08739-C02-01 (including FEDER). Pekka Hein\"am\"aki  was
supported by the Jenny and Antti Wihuri foundation.

\end{acknowledgements}

\end{document}